\documentclass[twocolumn,english,floatfix,prb,showpacs]{revtex4}
\usepackage[T1]{fontenc}
\usepackage[latin9]{inputenc}
\usepackage{color}
\usepackage{float}
\usepackage{amsmath}
\usepackage{graphicx}
\usepackage{amssymb}
\usepackage{esint}

\makeatletter
\@ifundefined{textcolor}{}
{%
 \definecolor{BLACK}{gray}{0}
 \definecolor{WHITE}{gray}{1}
 \definecolor{RED}{rgb}{1,0,0}
 \definecolor{GREEN}{rgb}{0,1,0}
 \definecolor{BLUE}{rgb}{0,0,1}
 \definecolor{CYAN}{cmyk}{1,0,0,0}
 \definecolor{MAGENTA}{cmyk}{0,1,0,0}
 \definecolor{YELLOW}{cmyk}{0,0,1,0}
 }

\@ifundefined{definecolor}
 {\usepackage{color}}{}
\@ifundefined{definecolor}
 {\@ifundefined{definecolor}
 {\usepackage{color}}{}
}{}

\usepackage{amsfonts}

\usepackage{amsfonts}

\usepackage{epstopdf}

\makeatother

\usepackage{babel}

\makeatother

\usepackage{babel}

\makeatother

\usepackage{babel}

\makeatother

\usepackage{babel}

\begin{document}

\title{Nearly frozen Coulomb liquids}

\author{Y. Pramudya, H. Terletska, S. Pankov, E. Manousakis, and V. Dobrosavljevi\'{c}}

\affiliation{Department of Physics and National High Magnetic Field Laboratory,
Florida State University, Tallahassee, Florida 32310}
\begin{abstract}
We show that very long-range repulsive interactions of a generalized
Coulomb-like form $V(R)\sim R^{-\alpha}$, with $\alpha<d$ ($d$-dimensionality),
typically introduce very strong frustration, resulting in extreme
fragility of the charge-ordered state. An {}''almost frozen''
liquid then survives in a broad dynamical range above the (very low)
melting temperature $T_{c}$ which is proportional to $\alpha$. This
{}''pseudogap'' phase is characterized by
unusual insulating-like, but very weakly temperature dependent transport,
similar to experimental findings in certain low carrier density systems. 
\end{abstract}


\maketitle

\section{introduction}

In designing novel materials, lightly doping a parent insulator is
typically the method of choice. An especially intriguing situation
is found in ultra-clean samples at finite doping, where neither the
Anderson \cite{anderson58} (disorder-driven) nor the Mott \cite{mott49}
(magnetism-driven) route for localization can straightforwardly succeed
in trapping the electrons. The tendency for charge ordering (CO) then
emerges as the dominant mechanism that limits the electronic mobility.
As first noted in early works by Wigner \cite{Wigner34} and Mott
\cite{mott49}, this is precisely where the incipient breakdown of
screening reveals the long-range nature of the Coulomb interactions.
The corresponding CO states proved to be of extraordinary fragility,
restricting the insulating behavior to extremely low densities and/or
temperatures \cite{tanatar89prl}. A broad range of parameters then
emerges where puzzling {}``bad insulator'' transport characterizes
such \emph{nearly-frozen Coulomb liquids}.

Unusual {}''bad-insulator'' transport behavior has been observed
in many systems. Examples range from high mobility two-dimensional
electron systems in semiconductors, \cite{huang:201302} to lightly-doped
cuprates, \cite{boebinger-prl96,panagopoulos-vlad04prb} manganites,
\cite{dagotto-2005-309} and even to the behavior of lodestone (magnetite)
above the Verwey transition. \cite{mott-book90} In all these cases,
a broad range of temperatures has been observed, where the resistivity
rises at low temperatures, but it does so with surprisingly weak temperature
dependence. In contrast to conventional insulators, where the familiar
activated transport reflects a gap for charge excitations, the {}``bad
insulator'' behavior has been interpreted \cite{mott-book90} as
a \emph{precursor} to charge ordering, leading to very gradual opening
of a soft pseudogap in the excitation spectrum.

The physical picture of \emph{a nearly-frozen Coulomb liquid} has
been proposed-on a heuristic level-by several authors, \cite{mott-book90,spivak01prb,andreev79}
providing a plausible and appealing interpretation of many experiments.
The interplay of spins and charge degrees of freedom
in pseudogap formation is still a controversial and unresolved problem.
Therefore, to focus on the corresponding role of charge fluctuations,
we deliberately ignore any spin effects, and consider a class of models
of spinless electrons interacting through long-range interactions.%

\begin{figure}[t]
\includegraphics[width=3.2in]{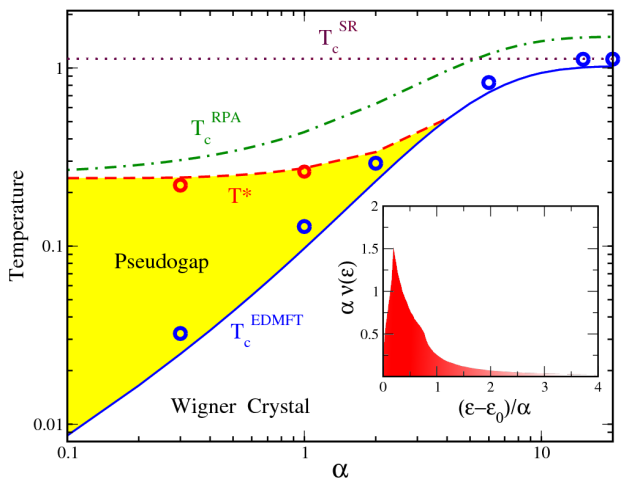}

\caption{(Color online) Phase diagram of the half-filled classical $d=3$ lattice
model with interactions $V(R)=R^{-\alpha}$. The charge ordering temperature
$T_{c}(\alpha)\sim\alpha$, as obtained from EDMFT theory (full line)
and Monte-Carlo simulations (open symbols). The pseudogap temperature
$T^{*}$(dashed line) remains finite as $\alpha\rightarrow0$; a broad
pseudogap phase emerges at $\alpha\le d$. We also show $T_{c}^{SR}\approx1$
for the same model with short-range interactions (dotted line), and
$T_{c}^{RPA}$ (dot-dashed line) from the classical limit of RPA.
The inset shows the corresponding plasmon mode spectral density, which
assumes a scaling form for $\alpha\ll1$. The fluctuations of these
very soft {}''sheer plasmons'' lead to the dramatic decrease of
$T_{c}$.}

\end{figure}

We present the simplest consistent theory of this strongly
coupled liquid state. We demonstrate that the existence
of such an intermediate liquid regime ,which emerges at $k_{B}T_{c}<k_{B}T\ll E_{c}$
(see below), is a very general phenomenon reflecting strong frustration
produced by long-range interactions. It holds for any interaction
of the form $V(R)\sim R^{-\alpha}$, both in continuum and lattice
models at any dimension $d\geq2$, with $\alpha\ll d$. Ours is a microscopic theory that substantiates this physical picture,
\cite{mott-book90,andreev79} based on quantitative and controlled
model calculations. We present a physically transparent analytical
description using extended dynamical mean-field theory (EDMFT) to
accurately describe the collective charge fluctuations, and benchmark
our result using Monte-Carlo MC simulations.

\section{our model and the edmft approach}

It has long been appreciated \cite{tanatar89prl,efros92,thakur-neilson} that in
Coulomb systems, the CO temperature scale $T_{c}$ is generally very
small as compared to the Coulomb energy $E_{c}=e^{2}/a$ ($a$ being
typical inter-particle spacing), which we use as our energy unit.
For example, for classical particles on a half-filled hypercubic lattice
$T_{c}\approx0.1$, \cite{efros92} while in the continuum and classical
Wigner crystal $T_{c}\approx0.01$ \cite{tanatar89prl}; similar results
are obtained both in $d=2$ and in $d=3$. Such large values of the
{}``Ramirez index'' \cite{ramirez94arms} $f=E_{c}/T_{c}$ suggest
that geometric frustration plays a significant role, reflecting the
long-range nature of the Coulomb force.

To clarify this behavior, we control the amount of frustration by
introducing generalized Coulomb interactions of the form $V(R)/E_{c}=(R/a)^{-\alpha}$.
We consider a lattice model of spinless electrons given by the Hamiltonian
\begin{equation}
H=-\sum_{ij}t_{ij}c_{i}^{\dagger}c_{j}+\frac{1}{2}\sum_{ij}V(R_{ij})(n_{i}-\langle n\rangle)(n_{j}-\langle n\rangle).\label{hamiltonian}\end{equation}
 Here $c_{i}^{\dagger}$ and $c_{i}$ are the electron creation and
annihilation operators, $t_{ij}$ are the hopping matrix elements,
$n_{i}=c_{i}^{\dagger}c_{i}$, and $R_{ij}$ is the distance between
lattice sites $i$ and $j$ expressed in the units of the lattice
spacing. The origin of frustration is then easily understood by noting
that in the classical limit our lattice gas model ($n_{i}=0,1$) maps
onto an Ising antiferromagnet ($S_{i}=\pm1$) with long-range interactions.
Here, the maximum level of frustration is achieved for infinite range
interactions ($\alpha\rightarrow0$), and any finite temperature ordering
is completely suppressed.

A controlled theoretical approach to our problem is available for
very long-range interactions ($\alpha\ll1$), which effectively corresponds
to a very large coordination number. In this limit the spatial correlations
assume a simplified form \begin{align}
 & G_{k}(i\omega_{n})=\left\langle c_{k}^{\dagger}c_{-k}\right\rangle =\frac{1}{i\omega_{n}-\epsilon_{k}-\Sigma(i\omega_{n})},\nonumber \\
 & \Pi_{k}(i\Omega_{n})=\left\langle n_{k}n_{-k}\right\rangle =\frac{\tilde{\Pi}(i\Omega_{n})}{\tilde{\Pi}(i\Omega_{n})+V_{k}},\label{sceq2}\end{align}
 where the momentum dependence of the (fermionic) self-energy $\Sigma(i\omega_{n})$
and the irreducible polarization operator $\tilde{\Pi}(i\Omega_{n})$
can be ignored %
\footnote{Strictly speaking, the nonlocal effects ignored by EDMFT are, for
$\alpha\ll1$, negligible throughout the broad pseudogap regime, but
not in the narrow critical regime close to $T_{c}$.%
}. A conserving approximation that formally sums all the corresponding
Feynman diagrams is given by the so-called EDMFT formulation, \cite{pastor-prl99,chitra00prl,pankov05prl}
where the relevant (local) quantities are computed from an auxiliary
local effective action\begin{multline}
S_{eff}=-\int d\tau d\tau^{\prime}c^{\dagger}(\tau)\mathcal{G}\mathrm{_{0}^{-1}}(\tau-\tau^{\prime})c(\tau^{\prime})\\
+\frac{1}{2}\int d\tau d\tau^{\prime}\delta n(\tau)\Pi_{0}^{-1}(\tau-\tau^{\prime})\delta n(\tau^{\prime}),\end{multline}
 where $\mathcal{G}\mathrm{_{0}^{-1}}(i\omega)=i\omega-\Delta(i\omega)$
and $\delta n(\tau)=n(\tau)-\langle n\rangle$. The dynamical effective-medium
(EM) functions $\Delta$ and $\Pi_{0}^{-1}$ represent the respective
fermionic and bosonic baths coupled to the given lattice site. For
a given bath, the (local) Dyson's equations stipulate that $\Sigma=\mathcal{G}\mathrm{_{0}^{-1}}-G_{\rm{loc}}^{-1}$
and $ \tilde{\Pi}^{-1}=\Pi_{\rm{loc}}^{-1}-\Pi_{o}^{-1}$, where $G_{\rm{loc}}$
and $\Pi_{\rm{loc}}$ are calculated directly from $S_{\rm{eff}.}$The self-consistency
loop is then closed by relating the local and the EM correlators,
viz, $G_{\rm{loc}}=\sum_{k}G_{k}(\Sigma)$ and $\Pi_{\rm{loc}}=\sum_{k}\Pi_{k}(\tilde{\Pi)}$.

\begin{figure}[t]
 \vspace{0.1cm}

\includegraphics[width=3in]{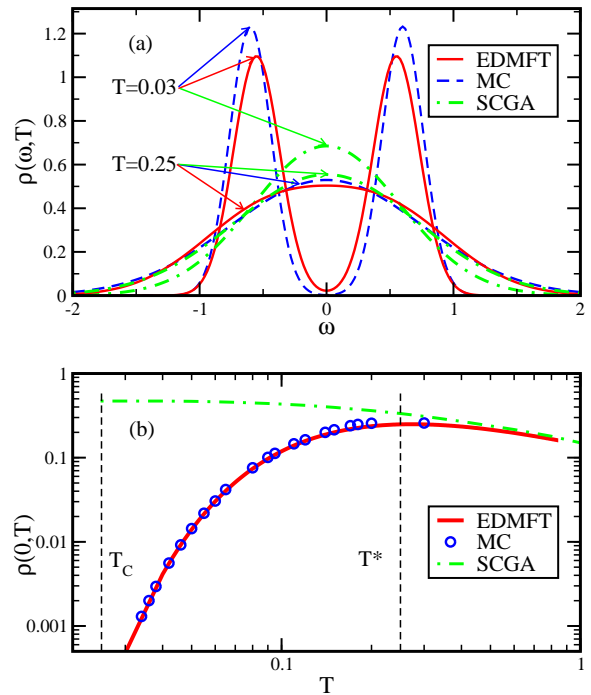}\caption{(Color online) (a) Density of states $\rho(\omega,T)$ obtained
with three different methods: EDMFT (full line), MC (dashed line)
and SCGA (dot-dashed line). Results are shown for the $d=3$ half-filled
cubic lattice with $t=0$, $a=0.3$, and two temperatures: $T=0.03\approx T_{c}$
and $T=0.25\approx T^{*}$. $\rho(\omega,T)$ obtained from EDMFT
(full line) agree well with MC results (dashed line), while SCGA (dot-dashed
line) does not account for the pseudogap formation. (b) Both EDMFT
(full line) and MC results (open symbols) $\rho(\omega=0,T)$ show
pseudogap opening (dramatic DOS decrease) at $T<T^{*}$ in contrast
to SCGA results (dot-dashed line). }

\end{figure}

\section{Classical limit }

The most stringent test for the accuracy of EDMFT is provided by examining
the classical limit ($t=0$), where pseudogap formation is most pronounced.
Here, the EDMFT equations can be solved in closed form, \cite{pankov05prl}
since the {}``memory kernel'' $\Pi_{0}^{-1}(\tau-\tau^{\prime})$
becomes a time-independent constant, $\Pi_{0}^{-1}=D/\beta^{2}$,
and the corresponding mode-coupling term in Eq. (3) can be decoupled
by a static Hubbard-Stratonovich transformation. The density correlator
then assumes the form $ $$\Pi_{k}=(4+D+\beta V_{k})^{-1},$ and the
self-consistency condition reduces to \begin{equation}
\frac{1}{4}=\int d\varepsilon\,\nu(\varepsilon)\left(4+D+\beta\varepsilon\right)^{-1},\label{scedmft}\end{equation}
 where we introduced the (classical) plasmon-mode spectral density
$\nu(\varepsilon)=\sum_{k}\delta(\varepsilon-V_{k}).$ The CO critical
temperature $T_{c}(\alpha)$ is identified by the vanishing of $\Pi_{k}^{-1}$
at the corresponding ordering wave vector $k=Q$ $ $. The mechanism
for $T_{c}$ depression is then easily understood by noting that for
$\alpha\ll1$ $ $the spectral density $\nu(\varepsilon)$ assumes
the scaling form $\nu(\varepsilon)=\alpha^{-1}\widetilde{\nu(}(\varepsilon-\varepsilon_{0})/\alpha)$,
where $\varepsilon_{0}\approx-1$; the explicit form of the scaling
function $ $$\widetilde{\nu}(\varepsilon-\varepsilon_{0})$ corresponding
to the half-filled cubic lattice is shown in the inset of Figure 1.
It features a sharp low-energy spectral peak of the usual dispersive
form $\nu(\varepsilon)\sim\varepsilon^{(d-2)/2}$ only at $(\varepsilon-\varepsilon_{o})<\varepsilon^{*}(\alpha)$,
i.e below a characteristic energy scale $\varepsilon^{*}(\alpha)\sim\alpha$
and a long high-energy tail of the form $\nu(\varepsilon)\sim\varepsilon^{-2}$.
Physically, these low energy excitations correspond to {}``sheer''
plasmon modes with wave vector$k\approx Q$; the scale $\varepsilon^{*}(\alpha)\sim\alpha$
thus plays a role of an effective Debye temperature. Its smallness
sets the scale for the ordering temperature $T_{c}(\alpha)=\alpha\int d\varepsilon\,\widetilde{\nu(}\varepsilon)/\epsilon\sim\epsilon^{*}(\alpha)$,
in agreement with an estimate based on a Lindemann criterion applied
to the sheer mode %
\footnote{The smallness of the melting temperature for a classical (continuum)
Wigner crystal can similarly be understood \cite{tanatar89prl} by
comparing it to the Debye temperature of sheer phonons.%
}.

In the classical limit, the single particle density of states (DOS)
$\rho(\omega,T)\equiv-\rm{Im}{G(\omega+i0^{+})}/\pi$ assumes a simple
bimodal form: \begin{multline}
\rho(\omega,T)=\frac{\beta}{\sqrt{8\pi D}}\left\{ \exp{\left[-\frac{\beta^{2}}{2D}\left(\omega+\frac{D}{2\beta}\right)^{2}\right]}\right.\\
\left.+\exp{\left[-\frac{\beta^{2}}{2D}\left(\omega-\frac{D}{2\beta}\right)^{2}\right]}\right\} ,\label{classdos}\end{multline}
 with the self-consistently determined parameter $D(T)$ setting the
scale of the Coulomb pseudogap ({}``plasma dip'') $E_{gap}=D/\beta,$
which starts to open at the crossover temperature $T^{*}=D/4\beta.$
We stress that, in contrast to the ordering temperature $T_{c}\sim\alpha,$
both $E_{gap}$ and $T^{*}$ remain finite for $\alpha\ll1,$ since
$D(T)\approx\beta$ in this limit. This leads to the emergence of
a broad pseudogap regime for $\alpha\lesssim d$, independent of the
precise form or the filling of the lattice. Remarkably, since $D(T)$
remains finite as $\alpha\rightarrow0$, both the density of states
$\rho(\omega,T)$ and the conductivity $\sigma(T)$ (see below) display
only very weak $\alpha$-dependence, in contrast to $T_{c}(\alpha)\sim\alpha$.

We benchmark these analytical predictions against
MC simulations which used careful finite-size
scaling analysis and (generalized) Ewald summation techniques to account
for long-range interactions (the detail is in the appendix). It was found that EDMFT captures all
qualitative and even quantitative features of the pseudogap regime
for several different values of the exponent $\alpha$, both in dimensions
$d=2$ and in $d=3.$ The detailed comparison of EDMFT and MC results
will be presented elsewhere; here we illustrate these findings for
a $d=3$ half-filled cubic lattice. Figure 1 shows how EDMFT accurately
captures the $\alpha$-dependence of $T_{c}$, which is found to decrease
in a roughly linear fashion as $\alpha\rightarrow0,$ while the $T^{*}\approx0.25$
remains finite, producing a large separation of energy scales and
a well-developed pseudogap regime. Note that the familiar Coulomb
interaction ($\alpha=1$) lies well within the small-$\alpha$ regime.
This observation makes it clear why our EDMFT theory remains very
accurate (as noted in previous work \cite{pankov05prl}) not only
for $\alpha\ll1$, but also for the physically relevant Coulomb case
$\alpha=1.$

\begin{figure}[t]
 \vspace{0.1cm}

\includegraphics[width=3.5in]{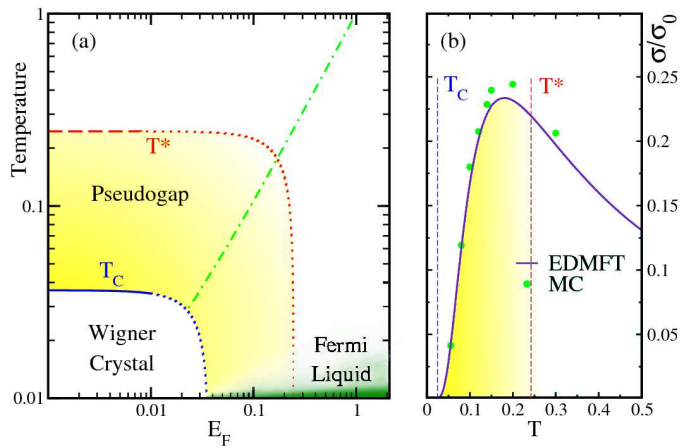}

\caption{(Color online) (a) EDMFT phase diagram for a half-filled cubic lattice
with $\alpha=0.3$ as a function of temperature $T$ and the electron's
Fermi energy $E_{F}\sim t$. Our semiclassical solution is valid
above the CO freezing temperature $T_{c}(E_{F})$ (full line), and
the Fermi liquid crossover temperature $T_{\rm{cros}}(E_{F})$ (dot-dashed
line). At intermediate temperatures $T_{c}<T<T^{*}$ we find well
developed pseudogap behavior, where transport assumes insulating-like
but very weak temperature dependence [as shown in (b)]; in the CO
phase ($T<T_{c})$ transport assumes the conventional activated form
(not shown). (b) Temperature dependence of the conductivity in the
semiclassical regime ( $E_{F}\ll1$), where only the prefactor $\sigma_{o}=\frac{\pi}{3}\frac{e^{2}t^{2}}{\hbar a}$
displays significant $t$-dependence; EDMFT results (full line) again
show remarkable agreement with results obtained by calculating $\rho(\varepsilon,\omega)$
in the classical limit using MC simulations (symbols). }

\end{figure}

\section{gaussian theories do not capture pseudogap formation\emph{ }}

The excellent comparison between EDMFT and MC results for the DOS
is shown for $\alpha=0.3$ in Figure 2(a). In contrast, the conventional
approaches \cite{schmalian_pines99prb}, which typically assume Gaussian
statistics for the collective charge fluctuations, fail to capture
the pseudogap opening at $T>T_{c}$. For example, the familiar self-consistent
Gaussian approximation ({}``spherical model''), while predicting
the exact same $T_{c}$ as EDMFT, produces Gaussian-shaped DOS at
any $T>T_{c}$, in contrast with MC findings; these shortcomings are
especially dramatic for $\alpha\ll d$ (see Figure 2). The popular {}``random-phase
approximation'' (RPA), \cite{lilly03prl} which amounts to a non-self-consistent
Gaussian approximation (SCGA), proves even less reliable in this regime.
It grossly overestimates the freezing temperature $T_{c}$, which
is found [dashed line in Figure 2 (b)] to remain finite even as $\alpha\rightarrow0$,
completely missing the pseudogap regime (shaded area in Figure 1). Physically,
the RPA (Stoner-like) freezing criterion reduces to the simplistic
Hartree (static mean-field) approximation, which ignores the dramatic
fluctuation effects of the soft collective (sheer plasmon) modes.

\section{bad-insulator transport in the semiclassical regime}

We expect the {}``bad insulator'' transport to be best pronounced
in the semiclassical regime $t\ll1,$ where the Coulomb energy represents
the largest energy scale in the problem. Here, the pseudogap phase
is reached by thermally melting the CO state at $T>T_{c}(t).$ While
our EDMFT equations are difficult to solve in general, in this incoherent
regime it is well justified to utilize an adiabatic ({}``static'')
approximation, \cite{schmalian_pines99prb} which ignores the time
dependence of the collective mode. The EDMFT equations can then be
solved in a manner similar to that in the strict classical limit (see
above), and we find\begin{align}
 & G(i\omega)=\int d\phi G_{\phi}(i\omega)P(\phi),\nonumber \\
 & \Pi(i\Omega)=T\sum_{i\omega}\int d\phi G_{\phi}(i\omega+i\Omega)G_{\phi}(i\omega)P(\phi),\nonumber \\
 & P(\phi)=\frac{1}{Z}\exp{\left(-\frac{D}{2}\phi(\phi+1)-\sum_{\omega}\ln{(G_{\phi}(i\omega))}\right)},\nonumber \\
 & G_{\phi}^{-1}(i\omega)=G_{0}^{-1}(i\omega)+\phi TD.\label{quasiclass}\end{align}
 Physically, the electrons travel in the presence of a static, but
spatially fluctuating random field representing the collective mode.
Its probability distribution $P(\phi)$ assumes a strongly non-Gaussian
character, reflecting the charge discreteness captured by EDMFT, but
ignored by conventional Gaussian theories such as RPA.

The semiclassical approximation remains valid \cite{schmalian_pines99prb}
as long as the time-dependence of the density correlator $\Pi(\tau)$
can be ignored, corresponding to \begin{equation}
|(\Pi(0)-\Pi(\beta/2))/\Pi(0)|\ll1.\label{validity}\end{equation}
 This criterion provides an estimate for the crossover temperature
$T_{cros}$, below which we expect (at large $t$) a gradual crossover
towards Fermi liquid behavior. The resulting phase diagram is shown
on Figure 3 (a).

To calculate transport, we use the Kubo formula for the resistivity,
which within the EDMFT theory assumes the form \cite{georgesDMFT1996}:\begin{equation}
\sigma=\frac{\pi}{3}\frac{e^{2}t^{2}}{\hbar a}\int_{-\infty}^{+\infty}d\varepsilon\int_{-\infty}^{+\infty}d\omega\rho_{o}(\varepsilon)\frac{A^{2}(\varepsilon,\omega)}{4T\cosh^{2}{\frac{\omega}{2T}}}\end{equation}
 where $\rho_{o}(\varepsilon)$ is the bare single-electron density
of states and $A(\varepsilon,\omega)=-\frac{1}{\pi}\rm{Im}{\left(\omega+i0^{+}-\varepsilon-\Sigma(\omega+i0^{+})\right)^{-1}}.$
In this adiabatic approximation, we calculate conductivity
in the leading order of $t^{2}$, in terms of quantities for $t=0$
($\rho(\varepsilon)$ and $A(\varepsilon,\omega)$).

These equations are easy to solve for arbitrary parameters of our
model, but we illustrate our findings in Figure 3, by showing explicit
results for half-filled cubic lattice with $\alpha=0.3$. Our semiclassical
solution is found to be valid in a broad pseudogap regime $T_{c}<T<T^{*}$,
which spans almost an order of magnitude in temperature (for $E_{F}\ll1$ we find
$T_{c}\approx0.03$ and $T^{*}\approx0.25$). Here the conductivity
displays unusual, insulating-like [$d\sigma(T)/dT>0$], but rather
weak (almost linear) temperature dependence [shown in Figure 3(b)],
surprisingly similar to that observed in magnetite above the Verwey
transition. Our microscopic theory confirms the heuristic picture
first proposed in early work of Mott. \cite{mott-book90}

\section{Conclusions}

We argued that pseudogap behavior in Coulomb systems directly reflects
strong frustration found in any system with very long-range repulsive
interactions. We demonstrated that a quantitatively accurate strong-coupling
description of this regime is possible using the interaction power
$\alpha$ as a small parameter in the theory. The corresponding EDMFT
equations were solved in the semiclassical regime where the pseudogap
phenomena are most pronounced, explaining {}``bad-insulator'' transport
found in many puzzling experiments. It should be noted that, using
appropriately formulated quantum impurity solvers, \cite{Werner06}
the same formulation could be extended to investigate low-temperature
quantum critical behavior for the same class of models. This fascinating
direction remains a challenge for future work.

\section{acknowledgement}

The authors thank Seng Cheong, Misha Fogler, Daniel Khomskii, Andy
Millis, Joerg Schmalian, Dan Tsui, and Kun Yang for useful discussions.
This work was supported by the National High Magnetic Field Laboratory
(YP, HT, SP, EM, and VD) and the NSF through Grants Nos. DMR-0542026 and
DMR-1005751 (Y.P., H.T., and V.D.).

\section{appendix}

\subsection{Ewald Potential\emph{ }}

In order to compute the effective potential of long-range interaction
$1/|\vec{r}_{ij}|^{\alpha}$ in hypercubic lattice \begin{eqnarray}
V(\vec{r}_{ij}) & = & \sum_{n\epsilon\mathbb{Z}}\frac{1}{|\vec{r}_{ij}+L\vec{n}|^{\alpha}},\label{eq:a.1}\end{eqnarray}
we use an Ewald-type summation \cite{ewald} with the help of the integral representation
of \cite{Ewald-JChem,ewald-smith} \begin{equation}
\frac{1}{|\vec{r}|^{\alpha}}=\frac{1}{\Gamma(\alpha/2)}\intop_{0}^{\varepsilon}t^{\frac{\alpha}{2}-1}e^{-r^{2}t}dt+\intop_{\varepsilon}^{\infty}t^{\frac{\alpha}{2}-1}e^{-r^{2}t}dt.\label{eq:a.2}\end{equation}
where $\Gamma(\alpha/2)$ is Gamma function. We switch the first term
of the integral to a momentum sum because the sum does not converge
rapidly in  real space. Next, we use the representation \cite{PhysRevB.75.144201}\begin{eqnarray}
 &  & \intop d^{d}r\sum_{\vec{n}}\delta(\vec{r}-[L\vec{n}+\vec{r}_{ij}])f(\vec{r})\nonumber \\
 & = & \int d^{d}r\sum_{\vec{G}_{l}}[e^{i\vec{G}_{l}.\vec{r}}-\delta(\vec{r})]f(\vec{r}),\label{eq:a.3}\end{eqnarray}
where $f(\vec{r})$ is any arbitrary function and on the right-hand
side the summation is over the vectors of the reciprocal lattice.
We then integrate $\vec{r}$ out and change the variable of the integration
in the first term $t\rightarrow1/t$. The final expression of the
potential takes the form\begin{eqnarray}
V(\overrightarrow{r}) & = & \frac{1}{\Gamma(\frac{\alpha}{2})}\sum_{\vec{n}}\varepsilon^{\alpha}\phi_{\frac{\alpha}{2}-1}(\varepsilon^{2}|\overrightarrow{r}+\overrightarrow{n}L|^{2})\nonumber \\
 & + & \frac{1}{\Gamma(\frac{\alpha}{2})\Omega}\sum_{\vec{k}\ne0}\pi^{\frac{3}{2}}\varepsilon^{\alpha-3}\phi_{\frac{1-\alpha}{2}}(\frac{|\vec{k}|^{2}}{4\varepsilon^{2}})e^{-i\vec{k}.\vec{r}}\nonumber \\
 & - & \frac{2\varepsilon^{\alpha}}{\Gamma(\frac{\alpha}{2})},\label{eq:a.4}\end{eqnarray}
where in each component vector $k_{i}=2\pi n_{i}$ and $n_{i}\epsilon\mathbb{Z}$.
At the maximum size of our Monte-Carlo simulation $L=24$, the potential
is accurate to the eighth decimal place with only $|n_{i}|=3$ in
each axis, and  $\epsilon=\sqrt{\pi}$ .

\subsection{Finite-size effects\emph{ }}

In the vicinity of Wigner crystallization, the finite-size effects
are very strong. The size dependence of the single particle density
of states  obtained from Monte-Carlo data for $d=3$, $\alpha=0.3$,
and $T=0.0554$ is shown in figure 4. To carry out a careful finite-size
scaling analysis of the DOS, we perform a two-Gaussian fit \begin{equation}
\rho(\omega)=h\left(e^{-\left((\omega-d)/w\right)^{2}}+e^{-\left((\omega+d)/w\right)^{2}}\right).\label{eq:b.1}\end{equation}
\begin{figure}[H]
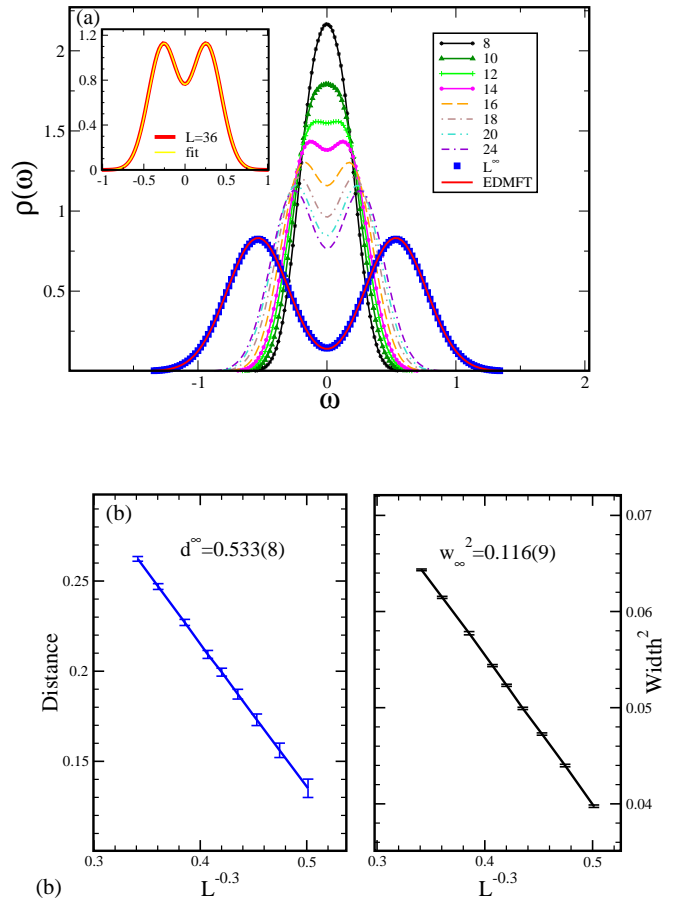

\bigskip{}

\includegraphics[scale=0.32]{4}

\medskip{}

\bigskip{}
\bigskip{}

\quad{}\includegraphics[scale=0.33]{5}\caption{(Color online) (a) Two-Gaussian fit of the single particle density
of states from different size of MC simulation. At this particular
temperature, They fit perfectly with the EDMFT results. The fit form
is shown in the inset. (b) The finite size scaling of the distance
(left) and width-squared (right) of the two-Gaussian function for different
sizes L =8,10,12,16,18,20,24. }

\end{figure}
The nonlinear (two-Gaussian) fitting is done using IGOR 6.01. The
fitting parameters, i.e., the distance between the Gaussian peaks and
width squared as a function of $L^{-\alpha}$ is shown in the  Fig 4(b). This allows us to perform an accurate extrapolation
to $L=\infty$, and the result is found to be in excellent agreement
with EDMFT prediction. Note how the finite-size result remains very
far from the $L=\infty$ extrapolant even for our largest system size
($L=24$). Accurate results, thus, simply cannot be obtained without
such finite size scaling analysis.


\end{document}